\documentstyle[12pt,times,psfig]{article}
\newcommand{\beq}{\begin{equation}}
\newcommand{\eeq}{\end{equation}}
\newcommand{\bea}{\begin{eqnarray}}
\newcommand{\eea}{\end{eqnarray}}

\begin{document}
\noindent{
{\large\bf 
LANL Report LA-UR-98-5999 (1998)\\
}
}
Talk given at the {\em Fourth Workshop on Simulating Accelerator Radiation
Environments (SARE4)}, \\
Knoxville, Tennessee, September 14-16, 1998

\begin{center}
{\Large \bf 
Improved Cascade-Exciton Model of Nuclear Reactions}\\
\vspace*{0.3cm}
{\bf Stepan~G.~Mashnik and Arnold~J.~Sierk}\\
\vspace{0.3cm}
{\it T-2, Theoretical Division, Los Alamos National Laboratory,
Los Alamos, NM 87545}\\
\end{center}
\begin{abstract}
Recent improvements to the Cascade-Exciton Model (CEM) of nuclear
reactions are briefly described. They concern mainly the cascade stage of
reactions and a better description of nuclei during the preequilibrium 
and evaporation stages of reactions. The development of the CEM 
concerning fission is given in a separate talk at this conference.
The increased accuracy and predictive power of the CEM are shown by 
several examples.  Possible further improvements to the CEM
and other models are discussed.  
\end{abstract}
\vspace{-0.3cm}
\begin{center}
{\large 1. Introduction} \\
\end{center}

During the last two decades, several versions of the Cascade-Exciton 
Model (CEM)~\cite{cem} of nuclear reactions have been developed
at JINR, Dubna.  A large variety of experimental data 
on reactions induced by nucleons \cite{cemnuclons}, pions \cite{cempions},
and photons \cite{cemphoto} has been analyzed in the framework of the CEM
and the general validity of this approach has been confirmed. 
Recently, the CEM has been extended by taking into account the competition
between particle emission and fission at the compound nucleus stage
and a more realistic calculation of nuclear level density~\cite{acta}. 
In the last few years, the CEM code has been modified~\cite{pit95,report97}  
to calculate 
hadron-induced spallation and used to study~\cite{pit95}--\cite{prokofiev} 
about 1000 reactions induced by nucleons from 10 MeV to 5 GeV 
on nuclei from Carbon to Uranium.

At present, several versions of the CEM code are used at a number of 
national laboratories and universities to solve different problems.
Several versions of the CEM code have been implemented 
with or without modifications
into a number of well-known transport codes, e.g., in:
MARS~\cite{mars98}, HETC~\cite{inhetc,hetc}, TIERCE~\cite{tierce},
MCNPX~\cite{inmcnpx}, and is currently being incorporated into
the LLNL radiation-transport code COG \cite{cog}.
The preequilibrium part of the CEM
has been implemented either as the initial Modified Exciton Model (MEM) 
code MODEX \cite{mem} or as the later version from \cite{cem95}
wholly or with some modifications into the tranport codes
HETC96~\cite{hetc96,hetc}, SHIELD~\cite{shield},
HETC-3STEP~\cite{hetc-3step,hetc}, HETC-FRG~\cite{hetc-frg,hetc},
CASCADE~\cite{cascade}, and into the intranuclear 
cascade-preequilibrium-evaporation-fission code INUCL~\cite{inucl}.

The recent
{\it International Code Comparison for Intermediate Energy Nuclear 
Data}~\cite{nea94a} has shown that the CEM adequately describes nuclear 
reactions at intermediate energies and has one of the best predictive
powers for double differential cross sections of secondary nucleons as
compared to other available models (see Tabs.~5 and 6 in the
Report~\cite{nea94a} and Fig.~7 in Ref.~\cite{cemphys}). 
As an example, in Fig.~1 we show experimental spectra of neutrons emitted
from interactions of 1.5 GeV protons with C, Al, Fe, In, and Pb
\cite{ishibashi97} compared with calculations using a version 
of the CEM as realized in the code CEM95~\cite{cem95}. We see that CEM95
reproduces the data very well.
In effect, these results can be regarded as a prediction of CEM95,
as the experimental data were published in 1997 while the calculations are
made without any changes of the code and using the previously determined
``preferred" set of input parameters (RM=1.5, IFAM=9, IDEL=0, ISH=6, ISHA=1,
IDELTA=1; see notations and details in the CEM95
code manual \cite{cem95}); these were fixed in 1995. 

\newpage

\begin{figure}[h!]
\vspace*{-3cm}
\centerline{
\psfig{figure=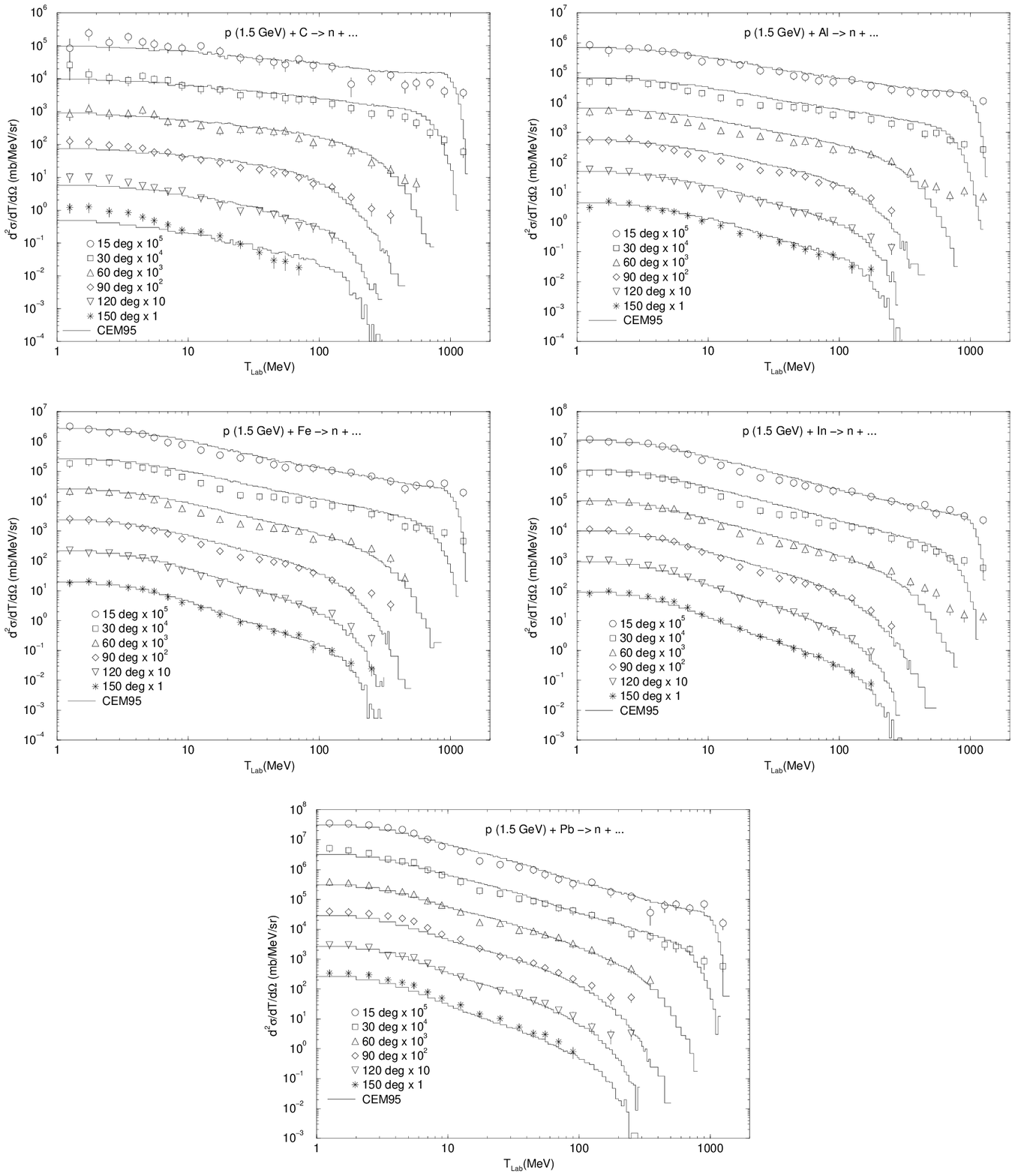,width=198mm,angle=0}}
\end{figure}
\vspace*{-3cm}

{\small
Fig.~1. 
Comparison of measured~\cite{ishibashi97}
double differential cross sections of neutrons from 1.5 GeV protons
on C, Al, Fe, In, and Pb with CEM95 calculations.\\
}

Unfortunately, not all CEM95 results agree so remarkably well with 
measuremetnts. As an example, Fig.~2 shows also experimental 
spectra of neutrons emitted from interactions of 256 MeV protons 
with Al~\cite{meier92}, Fe~\cite{meier92}, and Zr~\cite{stamer93}, 
and 1.5 GeV $\pi^+$ with Fe~\cite{nakamoto}. 
Calculations with CEM95 using the same 
  
\newpage

\begin{figure}[h!]
\vspace*{-3cm}
\centerline{
\psfig{figure=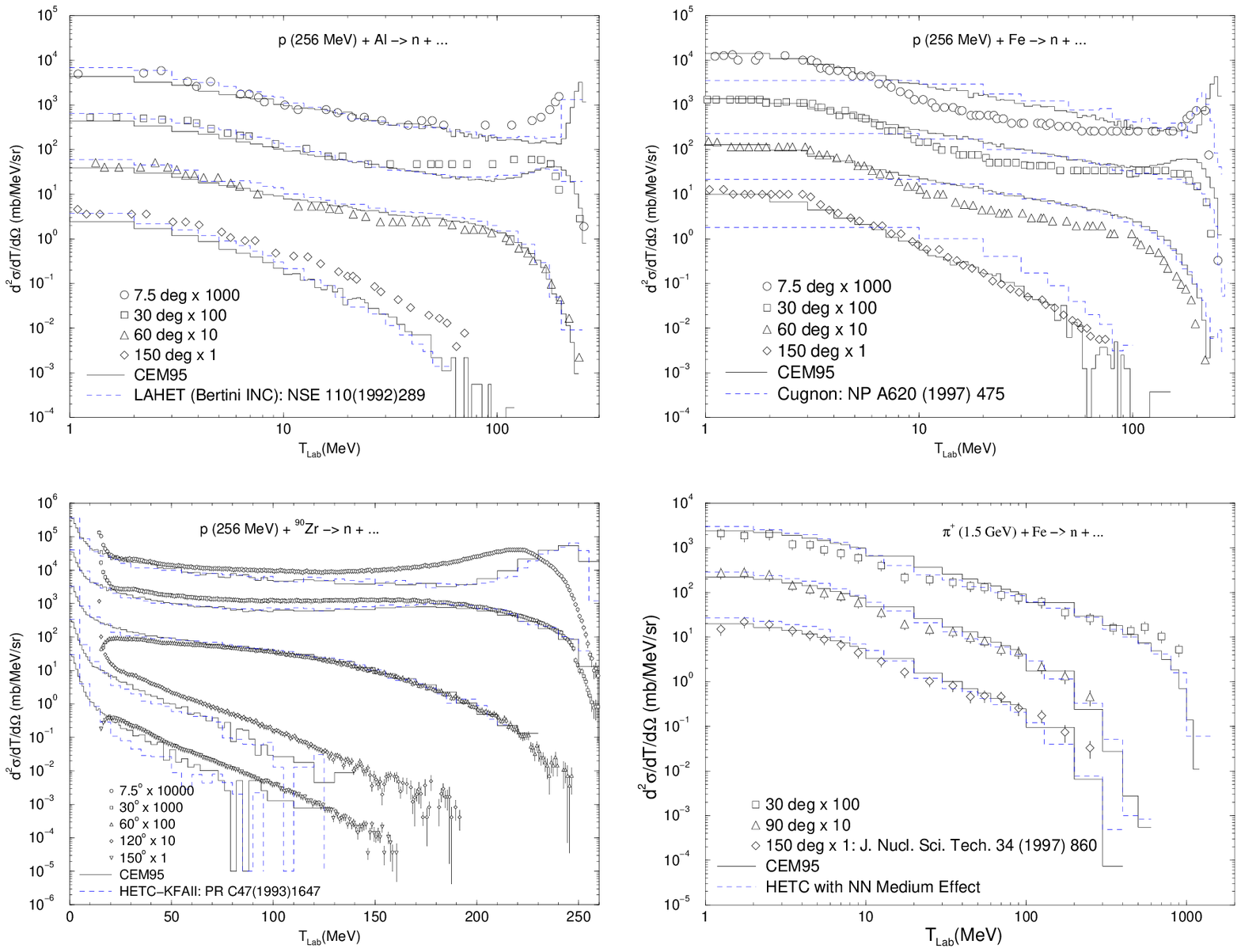,width=198mm,angle=0}}
\end{figure}
\vspace*{-9.5cm}

{\small
Fig.~2. 
Comparison of measured neutron spectra from 256 MeV protons
on Al \cite{meier92}, Fe \cite{meier92}, and Zr \cite{stamer93}, 
and 1.5 GeV $\pi^+$ on Fe \cite{nakamoto} with CEM95 calculations 
(solid histograms) and with results of LAHET \cite{lahet} 
for Al from \cite{meier92}, prediction of the recently
improved version of the Cugnon et al.~INC \cite{cugnon97} for Fe, 
HETC-KFAII \cite{hetc_kfa2,hetc} calculations for Zr from \cite{stamer93}, 
and a Japanese version \cite{modhetc} of HETC \cite{hetc} using 
modified NN cross sections 
to take account of the in-medium effects for 
$\pi^+$(1.5 GeV)+Fe from Ref.~\cite{nakamoto}, shown by dashed histograms.  
\\
}

\noindent{
fixed set of input parameters 
as for Fig.~1 are shown by solid histograms.
For comparison, results obtained 
with other well-known codes are shown as well:
LAHET \cite{lahet} calculations for Al from \cite{meier92}, 
a recent improved version of the Li\'{e}ge INC model by Cugnon et al. 
\cite{cugnon97} for Fe, a J\"{u}lich version of HETC \cite{hetc}
as realized in the code HETC-KFAII \cite{hetc_kfa2} 
for Zr from \cite{stamer93}, and a recent Japanese version 
\cite{modhetc} of HETC
using modified NN cross sections 
to take account of the in-medium effects for 
$\pi^+$(1.5 GeV)+Fe$\to$n+... from Ref. \cite{nakamoto}.
}

One can see that for these reactions the agreement of CEM95 results with
the measured neutron spectra is not as good as that shown in Fig.~1; 
in places the difference is bigger than a factor of 2. This seems to be
a characteristic of all the models compared here. 

For any model, to predict yields of isotopes produced at intermediate energies
is much more difficult than to calculate spectra of emitted particles \cite{paris97}. 
As one can see from Figs.~3 and 4 (adapted from
\cite{istc97}), CEM95 describes the majority of yields for isotopes
produced in the spallation region quite well and no worse than
other well-known codes, although for some individual isotopes the 
discrepancy with the

\newpage

\begin{figure}[h!]
\vspace*{-0cm}
\centerline{
\psfig{figure=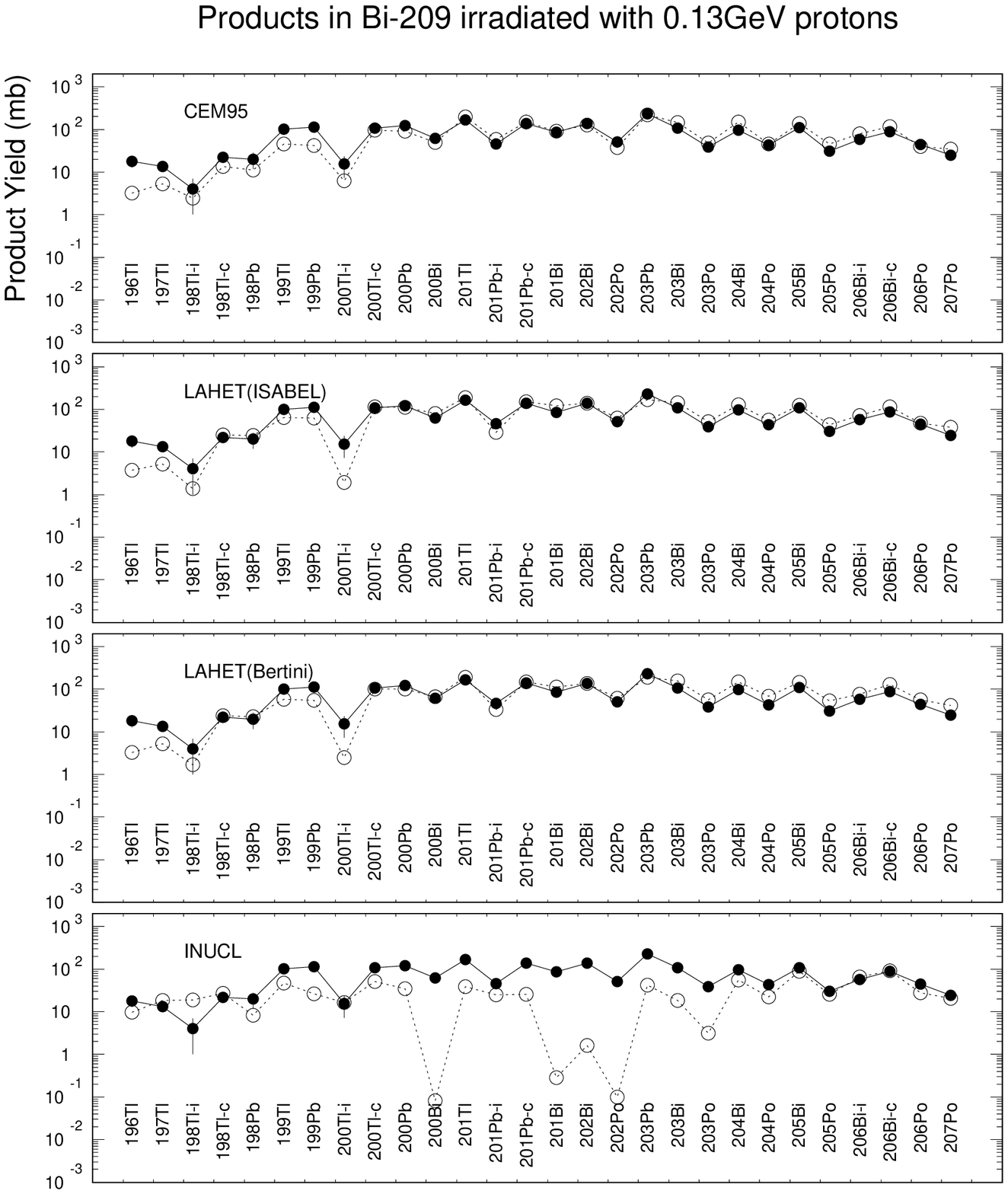,width=170mm,angle=0}}
\end{figure}
\vspace*{-1.5cm}

{\small
Fig.~3. 
Comparison between experimental data \cite{istc97} (filled circles)
and calculations with the codes CEM95 \cite{cem95}, LAHET \cite{lahet},
and INUCL \cite{inucl} (open circles) of yields of the indicated isotopes
from $^{209}$Bi irradiated with 130 MeV protons.
\\
}

\newpage

\begin{figure}[h!]
\vspace*{-0cm}
\centerline{
\psfig{figure=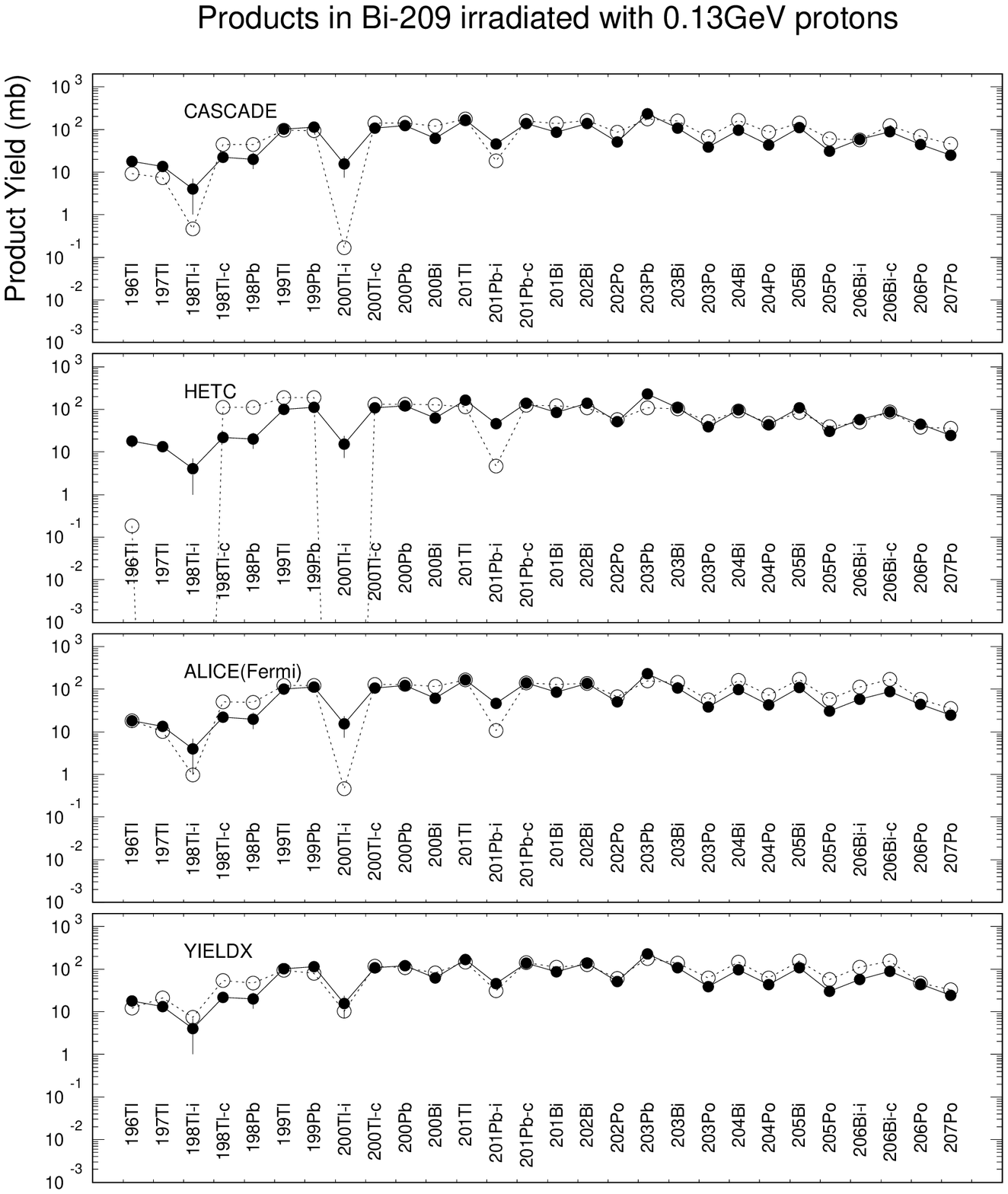,width=170mm,angle=0}}
\end{figure}
\vspace*{-1.5cm}

{\small
Fig.~4. 
The same as Fig.~3 using instead calculations with the
CASCADE \cite{cascade}, HETC \cite{hetc}, HMS-ALICE \cite{alice96} codes,
and YIELDX phenomenological systematics \cite{yieldx}.
\\
}
\newpage
\noindent{ cross sections. 
measurements is quite large.

The recent projects for Accelerator Transmutation of Waste (ATW) and 
Accelerator Production of Tritium (APT) have revived interest in accurate 
nuclear reaction data at intermediate energies (see, e.g., ~\cite{kalmar96}).
Not all needed data can be measured, therefore reliable models are 
required to provide the necessary 
This is our motivation for attempting to improve the CEM into a model capable 
of predicting reliable nuclear cross sections for arbitrary targets in a wide 
range of incident energies.}

The philosophy of our work is not just to include a few more parameters,
fitting them to describe the available data, a commonly used practice 
in development of codes for applications. Instead,
we are further developing the CEM by progressively incorporating
features of previously neglected physics, striving meanwhile not to destroy
the current broad strengths of CEM95 and its good predictive power.
To investigate the dependence of theoretical results on the physics
incorporated in CEM95 and to identify the improvements to the CEM 
and to other similar models which are of highest priority,
we have performed a detailed analysis of more than 600 excitation
functions for proton-induced reactions on 19 targets from C to Au at
energies from 10 MeV to 5 GeV \cite{report97}. After this analysis, we
created a list of potential improvements 
to the CEM95 code \cite{report97}. Some these have already been 
incorparated and we describe them briefly in the present paper. 
Nevertheless, as we have not yet completed our work, some 
of the results presented below are still preliminarily.  

\begin{center}
{\large 2. The Main Concepts of the Model} \\
\end{center}

A detailed description of the CEM may be found in Ref.~\cite{cem}
and of its extended version, as realized in the code CEM95, in 
Refs.~\cite{pit95,report97};
therefore, we mention here only its basic assumptions and 
features modified in the present work, for the sake of clarity. The CEM 
assumes that reactions occur in three stages. The first stage is 
the intranuclear cascade in which primary and
secondary particles can be rescattered
several times prior to absorption by, or escape from the nucleus. 
The cascade stage of the interaction is described by the standard
version of the Dubna intranuclear cascade model (INC)~\cite{book}.
The excited residual nucleus remaining after the emission of the 
cascade particles determines the particle-hole configuration that is 
the starting point for the second, preequilibrium stage of the 
reaction. The subsequent relaxation of the nuclear excitation is 
treated by an extension of the Modified Exciton model 
(MEM)~\cite{mem} of preequilibrium decay which also
includes a description of the equilibrium evaporative third stage of 
the reaction.

All the cascade calculations are carried out in a three-dimensional 
geometry. The nuclear matter density $\rho(r)$ is described by a 
Fermi distribution with two parameters taken from the analysis of 
electron-nucleus scattering, namely
\begin{equation}
\rho(r) = \rho_p(r) + \rho_n(r) = \rho_0 \{ 1 +                      
exp [(r-c) / a] \} \mbox{ ,} 
\label{a1}
\end{equation}
where $c = 1.07 A^{1/3}$ fm, $A$ is the mass number of the target, and
$a = 0.545$ fm.  For simplicity, the target nucleus is divided by 
concentric spheres into seven zones in which the nuclear density 
is considered to be constant.
The energy spectrum of the target nucleons is 
estimated in the perfect Fermi gas approximation with the local Fermi 
energy
$T_F(r) = \hbar^2 [3\pi^2 \rho(r)]^{2/3}/(2m_N)$, where $m_N$ is the nucleon
mass.
The influence of intranuclear nucleons on the incoming projectile is
taken into account by adding to its laboratory kinetic
energy an effective real potential $V$, as well as by considering
the Pauli principle which forbids a number of intranuclear collisions
and effectively increases the mean free path of cascade particles inside
the target.
For incident nucleons
$V \equiv V_N (r) = T_F(r) + \epsilon$,
 where $T_F(r)$ is the corresponding Fermi
energy and $\epsilon$ is the mean binding energy of the nucleons 
($\epsilon \simeq 7$ MeV~\cite{book}).
For pions, in the Dubna INC one usually 
uses~\cite{book} a square-well nuclear potential 
with the depth $V_{\pi} \simeq 25$ MeV, independently of the nucleus and
pion energy.
The interaction of the incident particle with the nucleus is approximated as
a series of successive quasifree collisions of the fast cascade
particles ($N$ or $\pi$) with intranuclear nucleons:
\begin{eqnarray}
NN \to NN , \qquad NN \to \pi NN , \qquad  NN \to \pi _1,\cdots,\pi _i NN 
\mbox{ ,}                                                                 
\nonumber \\
\pi N \to \pi N, \qquad  \pi N \to \pi _1,\cdots,\pi _i N  \qquad  (i \geq 2) 
\ .
\label{a2}
\end{eqnarray}
\noindent{
To describe these elementary collisions, we use experimental cross
sections for the free $N N$ and $\pi N$  interactions, simulating 
angular and momentum distributions of secondary particles by
special 
polynomial expressions with energy-dependent coefficients,
and take into account the Pauli principle.
}

Besides the elementary processes (2), the Dubna INC also takes into account 
pion absorption on nucleon pairs 
\begin{equation}
\pi + [NN] \to NN .                                                    
\label{a3}
\end{equation}
The momenta of two nucleons participating in the absorption are chosen
randomly from the Fermi distribution, and the pion energy is distributed 
equally between these nucleons in the center-of-mass system
of the pion and nucleons participating in the absorption. The direction 
of motion of the resultant nucleons in this system is taken as
isotropically distributed in space.  The effective cross section for
absorption (let us speak below, for concreteness, e.g., about 
$\pi^-$) is estimated from the experimental cross-section 
for pion absorption by deuterons
\begin{equation}
\sigma ( \pi^- + ``np" \to nn) = W \cdot \sigma (\pi^- + d \to nn) \ .
\end{equation}
The quantity $W$ can depend on the pion energy
$T_{\pi}$, on the characteristics of the  
target nucleus, the point where the pion is absorbed, and on the
spin-isospin states of absorbing pairs
(see details and references in~\cite{cempions}). 

\begin{center}
{\large 3. Study Results} \\
\end{center}

{\bf A. Elementary Cross-Sections.}   
In the Dubna INC model~\cite{book} used in CEM95,
the cross sections for the free $NN$ and $\pi N$ interactions (2)
are approximated using a special algorithm of interpolation/extrapolation
through a number of picked points, mapping as well as possible the
experimental data.
This was done very accurately by the group of Prof.~Barashenkov using all
experimental data available at that time, about 30 years ago.
Currently, the experimental data on cross section is much 
richer, therefore we decided to revise the approximations of all elementary
cross sections used in CEM95. We started with collecting all published 
experimental data from all available sources.
Then, we developed an improved, as compared with the standard 
Dubna INC \cite{book},
algorithm for approximation of cross sections and developed simple
and fast approximations for elementary cross sections which fit very well
presently available experimental data not only to 5 GeV, the upper
recommended energy for the present version of the CEM, but up  to 50--100
GeV and higher, depending on availability of data. So far, we have such
approximations for 34 different types of elementary cross sections
induced by nucleons, pions, and gammas. Cross sections for other
types of interactions taken into account in the CEM are calculated from
isospin considerations using the former as input.

We consider this part of our CEM improvement as an independently useful
development, as our approximations are reliable, fast, and easy to
incorporate into any transport, INC, BUU, or Glauber-type model codes.
For example, our new approximations recently have been successfully 
incorporated by N.~Mokhov into the recent improved version of the 
MARS code system at Fermilab \cite{mars98}. 

\newpage

\begin{figure}[h!]
\vspace*{-0.2cm}
\centerline{
\psfig{figure=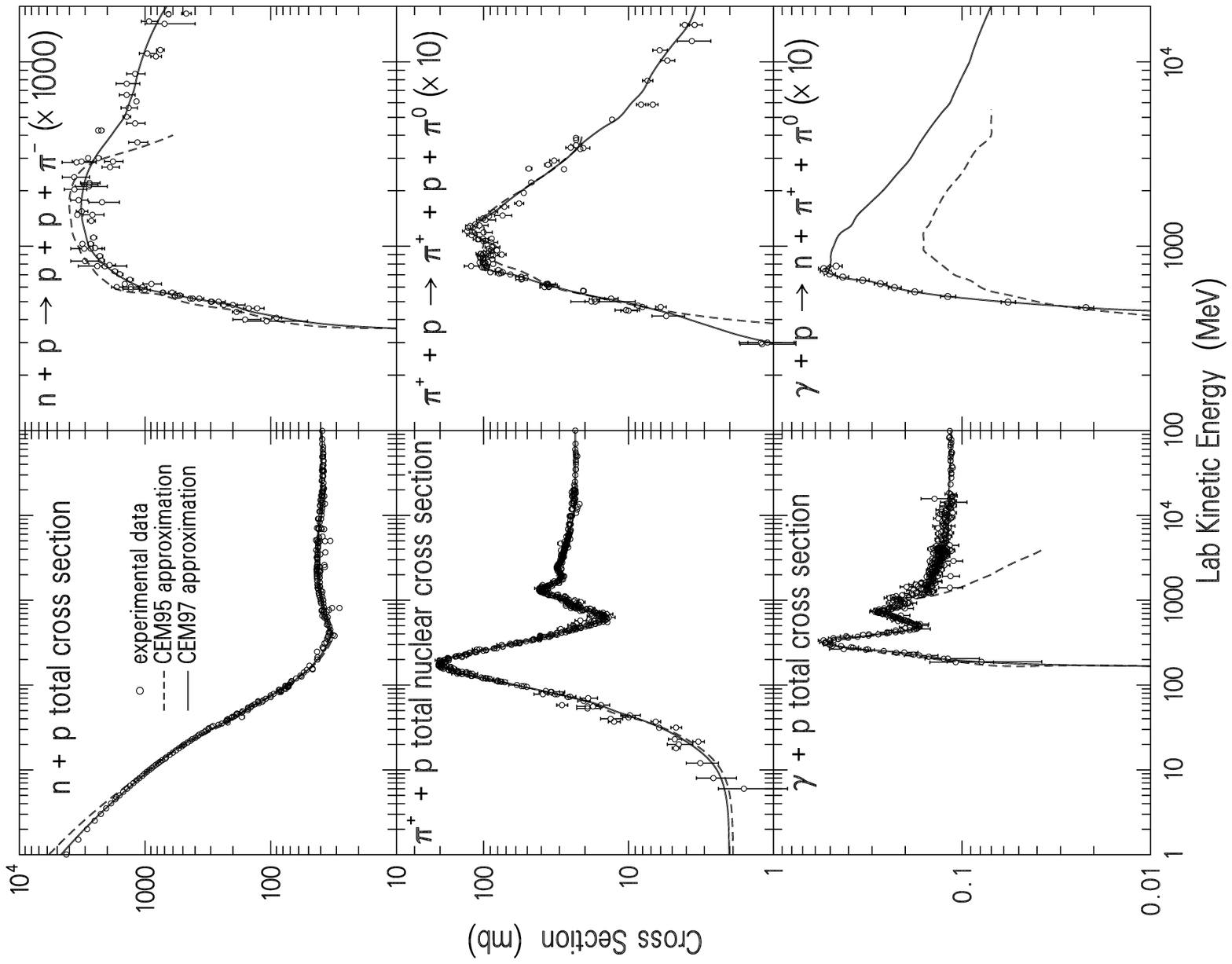,width=270mm,angle=-90}}
\end{figure}
\vspace*{+0.8cm}

{\small
{\bf Fig.~5}.
Energy dependence of the  $n p$, $\pi^+ p$, and $\gamma p$ total cross
 sections and of the $n p \to p p \pi^-$, $\pi^+ p \to \pi^+ p \pi^0$, and
$\gamma p \to n \pi^+ \pi^0$ ones.
Experimental points are from our compilation \cite{crossec}.
Solid lines are results of our present approximations; dashed lines
show the standard Dubna INC approximations \cite{book} used in CEM95. 
}

We are going to publish our compilation
of experimental data and our approximations of these cross sections
in a separate paper and to make them available to users through the Web 
\cite{crossec}, as well as to make available to users our code 
for these approximations. An example of 6
compiled experimental cross sections together with our new approximations
and the old approximations from CEM95 is shown in Fig.~5. We see that
our new approximations describe indeed very well all data.
Although presently we have much more data than 30 years ago when 
Barashenkov's group produced their approximations used in CEM95, for 
a number of interaction modes like the total cross sections shown 
in the left panel of Fig.~5,
 the original approximations also agree very well with
presently available data, in the energy region where the Dubna INC was 
developed to work.  This is a partial explanation of why the old
Dubna INC \cite{book} and the younger CEM95 \cite{cem95} work so well
for the majority of characteristics of nuclear reactions. 

On the other hand, for some modes of elementary interactions 
like the ones shown in the right panel of Fig.~5, the old approximations 
differ significantly from the present data, demonstrating the need for our 
present improvements for a better description of all modes of nuclear 
reactions.\\

{\bf B. Effects of Refraction and Reflection}.
In CEM95, the kinetic energy of
cascade particles is increased or decreased as they move from one 
potential region (zone) to another, but their directions remain unchanged.
That is, in our calculations, refraction or reflection of cascade particles at
potential boundaries is neglected.

To understand how our results depend on effects of refractions and reflections 
we performed calculations for a number of reactions with a modified version
of CEM95 taking into account refractions and reflections as described in 
the monograph \cite{book} and realized by A.~S.~Iljinov \cite{iljinov70}.
An example of our results is shown on Fig.~6. 
From these and other similar results obtained for other reactions 
we conclude that refractions and reflections affect only weakly
the spectra of secondary particles and somewhere more the total
inelastic/elastic cross sections. As one could expect in advance, 
at low incident energies, refractions and reflections
more strongly affect the calculated spectra but on the whole, CEM95 
results without taking into account refractions and reflections agree better 
with the  data, for the majority of analyzed reactions. Therefore, we 
prefer for our new version of the CEM to ignore refractions and 
reflections, as in the standard Dubna INC.
We note that similar results were obtained earlier by 
Chen et al.~\cite{chen68}. \\

{\bf C. Pion-Nucleus Potential $V_\pi$}.
It is natural that theoretical characteristics of a nuclear reaction
involving pions depend on the pion-nucleus potential used in calculations. 
This problem has a history of almost half a century; it is 
not solved completely even now, and its review is beyond the scope of the
present paper. 

As mentioned above, CEM95 uses a constant attractive pion-nucleus 
potential of 25 MeV.  This value was obtained 30 years ago by Barashenkov, 
Gudima, and Toneev \cite{barashenkov69} from an analysis of available
data and was suggested as a basis for the Dubna INC.
At the same time, analyzing pion-nucleus reactions at energies
below 300 MeV and nuclear absorption of stopped antiprotons 
with the Dubna INC, Iljinov, Nazaruk, and Chigrinov suggested~\cite{nazaruk} 
that one can achieve a better agreement with the data using $V_\pi = 0$.
Making the situation more confusing, in a recent
extension of the Dubna INC for photonuclear reactions at energies up to
10 GeV by Iljinov et al.~\cite{iljinov97}, a better description of pion
photoproduction was achieved for $V_\pi = 35$ MeV. 

Since previous INC studies of the values of $V_\pi$  are based
on analysis of quite scanty and old data, we decided to 
return once more to this question using the recent measurements at 
LAMPF of non-charge exchange (NCX) \cite{zumbro93},
single charge exchange (SCX) \cite{hoibrten,ouyang}, and 
double charge exchange (DCX) \cite{wood92}
pion production on targets from C to Bi in the enegy region
from 120 to 500 MeV.

\newpage

\begin{figure}[h!]
\vspace*{-3cm}
\centerline{
\psfig{figure=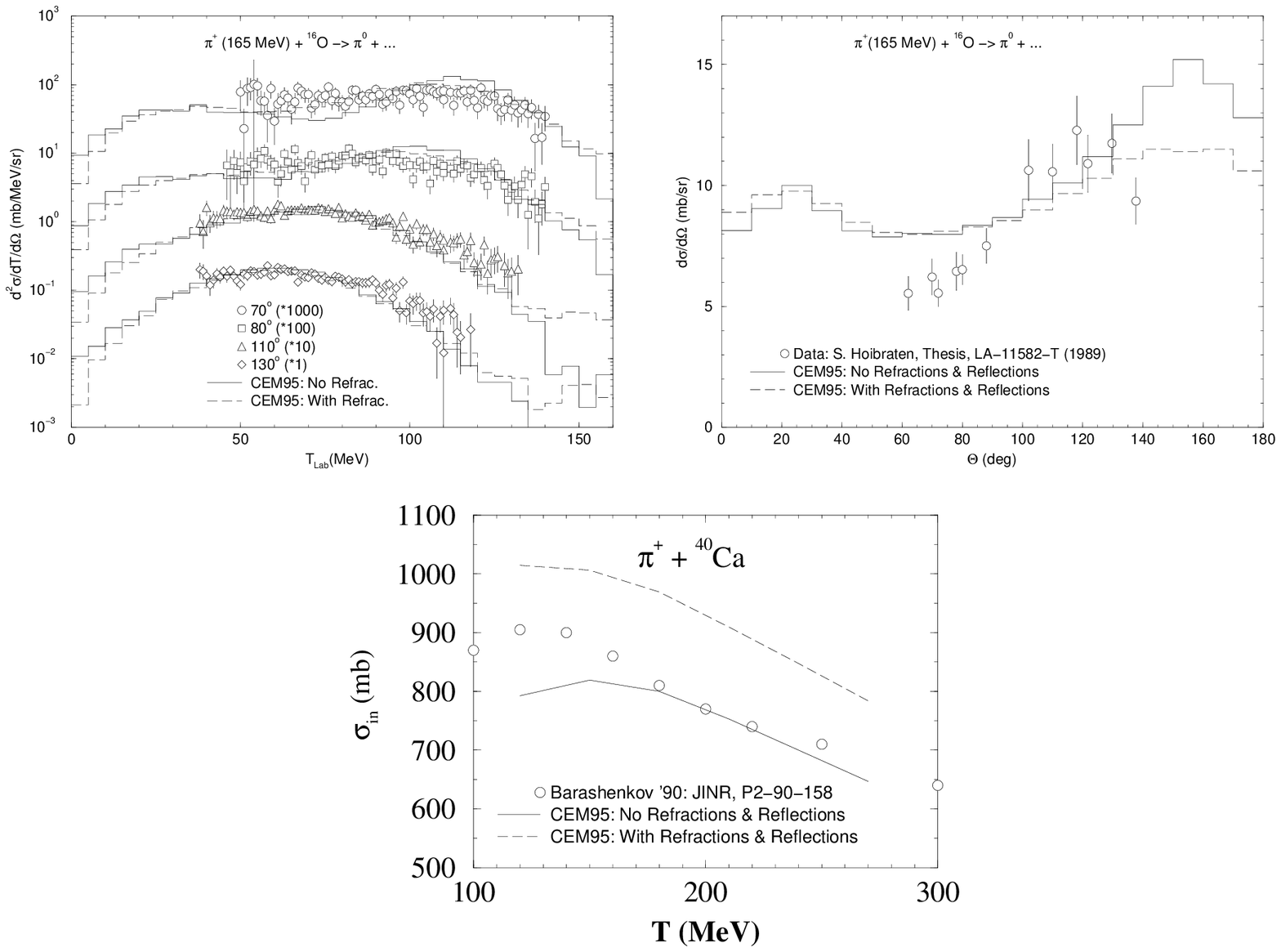,width=198mm,angle=0}}
\end{figure}
\vspace*{-10.0cm}

{\small
{\bf Fig.~6}.
Measured double differential spectra and angular distribution of $\pi^0$
from $\pi^+$(165 MeV)+$^{16}$O \cite{hoibrten} and Barashenkov's
systematics \cite{barashenkov90} for the
experimental $\pi^+$+Ca total inelastic cross section compared
with CEM95 calculations without (solid) and
with refractions and reflections (dashed). 
\\
}

We perform calculations for a large number of NCX, SCX, and DCX 
reactions using several values for $V_\pi$.  An example of our results 
is shown in Fig.~7. Our analysis shows that total inelastic cross sections 
and mean multiplicities of secondary pions, as well as NCX and SCX 
pion spectra at incident energies well above the $\Delta (1232)$ 
resonance region depend only slightly on the value of $V_{\pi}$.
But at lower incident energies corresponding to the $\Delta (1232)$ region
the shapes of all NCX, SCX, and DCX spectra change significantly 
with changes in the value of $V_{\pi}$. Unfortunately, none of the
fixed values of the $V_{\pi}$ we try can provide a good simultaneous
description of all data. Nevertheless, the best overall agreement 
with the
different data we analyse is achieved using the original value of
$V_{\pi}$ = 25 MeV.

Other recent studies show that $V_\pi$ is a complex function of 
pion energy $T_\pi$, nuclear target ($A,Z$), radius vector $r$, and on
the isospin of the pion.  Ideally, for any INC model it would be useful
to have a simple approximation for $V_\pi(A,Z,T_\pi,r)$ that could be
used in simulations.  We do not know any simple systematics for $V_\pi$, 
and it remains still an input parameter.
For the moment, we choose $V_\pi = 25$ MeV, with the understanding
that this is not a final solution to the problem. \\

{\bf D. Pion Transparency}.
Transparency means an inhibition of the interaction of a pion with
subsequent nucleons in the target nucleus for a certain time after 
creation of the pion. Recently, attention was 
called 
\\


\begin{figure}[h!]
\vspace*{-3cm}
\centerline{
\psfig{figure=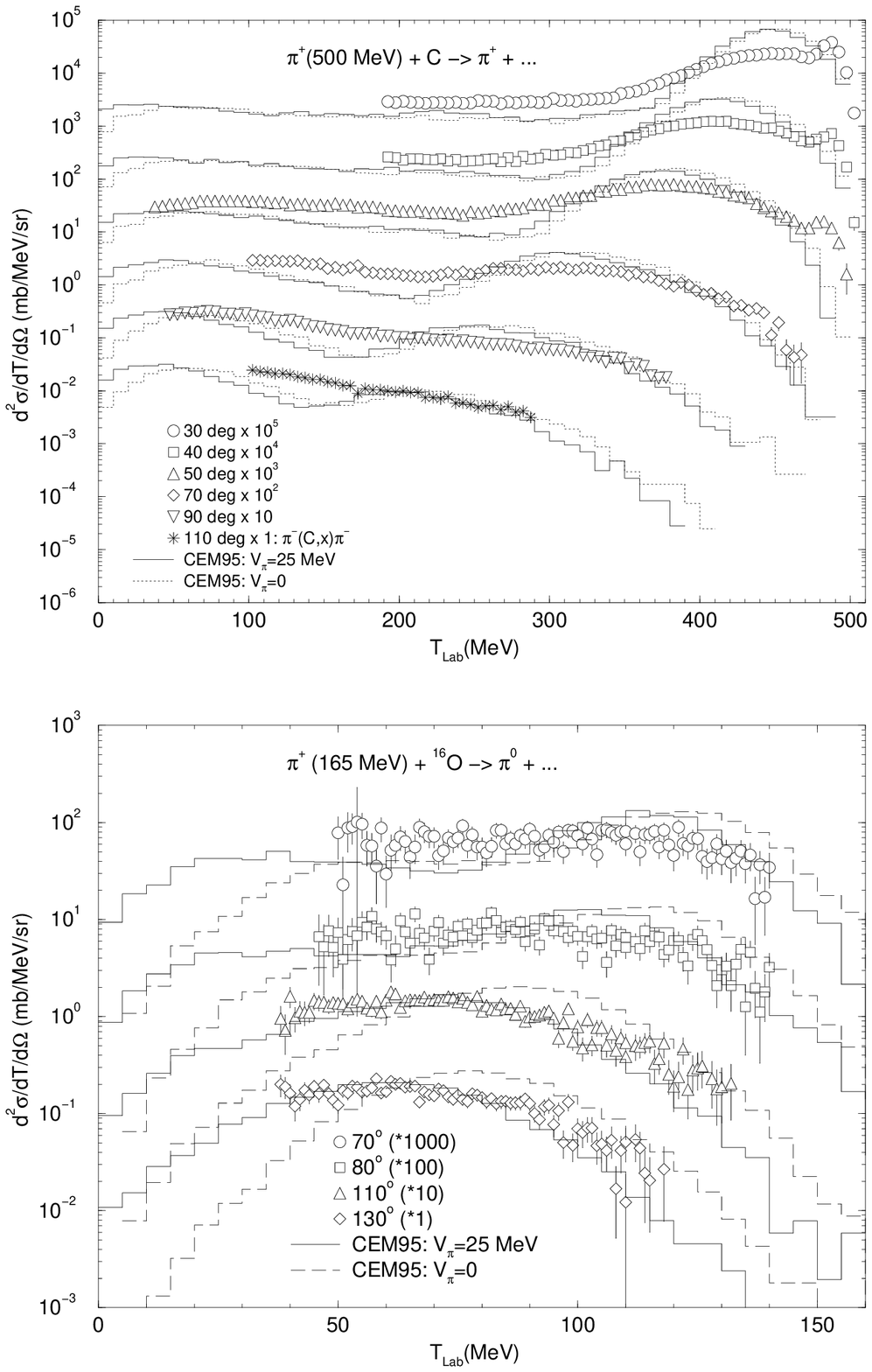,width=190mm,angle=0}}
\end{figure}
\vspace*{-3.5cm}

{\small
{\bf Fig.~7}.
Comparison of experimental
NCX \cite{zumbro93} and SCX \cite{hoibrten} pion spectra for
interactions of 500 MeV $\pi^+$ with C and 165 MeV $\pi^+$ with O
with standard CEM95 results 
($V_{\pi} = 25$ MeV; solid histograms)
and with calculations using
$V_{\pi} = 0$ MeV (dashed histograms).
\\
}

\noindent{
to this possible new 
physical effect after a study of the previously mentioned NCX pion 
production measurements at LAMPF \cite{zumbro93} with the INC model 
by Gibbs et al.~(GINC) \cite{gibbsinc}. 
}
GINC failed to reproduce the data, 
strongly underpredicting the measured yields at pion energies near 200 MeV.
When pions 
involved in $(\pi,2\pi)$ reactions were not allowed to interact 
with the nucleus for a time equal to 2 fm/c, a considerable improvement 
in the agreement with data was achieved \cite{zumbro93}.
This fact was interpreted as a new phenomenon of nuclear pion transparency 
with several possible exotic sources like: a need for a hadronization time
 for the wave function of pions involved in pion production to settle into 
an eigenstate,
a mechanism which allows the pions to be transported through the nucleus
by weakly interacting $\sigma$ mesons, 
or renormalized particle masses at central nuclear density ~\cite{zumbro93}.
Recently, we analyzed~\cite{colorado97} SCX $\pi^0$ spectra
from interactions of negative pions with $^{12}$C and $^7$Li at
467 and 500 MeV using the same GINC~\cite{gibbsinc} and 
we got results very similar to the ones on NCX spectra published by
Zumbro et al.~\cite{zumbro93}; we saw also an indication of some pion 
transparency.

Therefore we decided to implement this new piece of physics in the CEM,
if appropriate. The Dubna INC used in CEM95 does not take into account
the time of intranuclear elementary interactions, instead, it considers
the coordinates of these interactions. So, instead of a hadronization
time, we can require that the mean free path of secondary pions produced
in ($\pi,2\pi$) reactions to be longer than a fixed value, e.g., 2 fm. 
Including a transparency distance of 2 fm for pions from  $(\pi,2\pi)$
interactions into the CEM improves the description of
some of the NCX, SCX, and DCX pion spectra at pion incident energies
around the $\Delta(1232)$ resonance region. An even better agreement with
some of the data \cite{zumbro93}--\cite{wood92} is achieved by
imposing a transparency of 2 fm for pions created in a ($\pi,2\pi$) 
interaction and, in addition, a transparency of 1.2 fm for all other pions 
involved in a reaction.

As we mentioned in the beginning, the aim of our work is to incorporate
progressively in CEM95 new physical ingredients without destroying its
present wholeness and good predictive power. Therefore we consider it
necessary to analyze not only pion-induced NCX, SCX, and DCX reactions,
but also other types of intermediate energy nuclear reactions where
pion production, absorption, and emission are important channels.
We study a large variety of characteristics of different types of nuclear 
reactions from a unique point of view, without fitting any parameters 
except for several versions of pion transparencies.
Two  results of this study, one ``good" and one ``bad", are
shown in Fig.~8. The general conclusion we drawn from our analysis
is that imposing some transparency for pions in CEM95 improves
description of parts of the NCX, SCX, and DCX pion spectra 
around the $\Delta(1232)$ resonance region (see the upper plot in
Fig.~8) as well as of the 100--200 MeV region of
backward nucleon spectra from nucleon- and pion-induced
reactions. At the same time, incorporation of such  pion transparency 
destroys the self-consistency of the CEM95
and results in a clear overestimation of pion production at forward
angles (see the lower plot in Fig.~8) for different reactions. 
We interpret these results as an indication that we need a further 
improvement of the treatment of pion-nucleus interactions
in CEM95 (e.g., inclusion of $\Delta$ and other resonances explicitly
as cascade participants, a better approximation of experimental
$\pi N$ and $NN$ angular distributions and of pion absorption in 
the CEM) rather than as evidence of new physics related with some 
exotic pion transparency.  \\

{\bf E. Pion Absorption}.
The pion absorption cross section on nuclear pairs is treated in CEM95
according to Eq. (4), using experimental data for absorption on deuterons.
Calculations by many authors reveal that an overall satisfactory 
description of different experimental data can be obtained 
over a large range of pion energies and target nuclei,
provided that the ``effective" approximation $W \simeq const = 4$ is used.
The probability of absorption of pions on different $pp$, $np$, and $nn$
pairs is taken to be the same, whenever the absorption is allowed by the charge
conservation law. This is quite a rough approximation and we have encountered
problems using it to study pion-nucleus interactions in the $\Delta$
resonance region \cite{cempions}.

If we assume that pions are absorbed on nucleon pairs only through the
production with subsequent absorption of $\Delta$ isobars,
$\pi [N N] \to \Delta N \to N N$, than from the isospin decomposition
one gets the

\newpage

\begin{figure}[h!]
\vspace*{0.8cm}
\centerline{
\psfig{figure=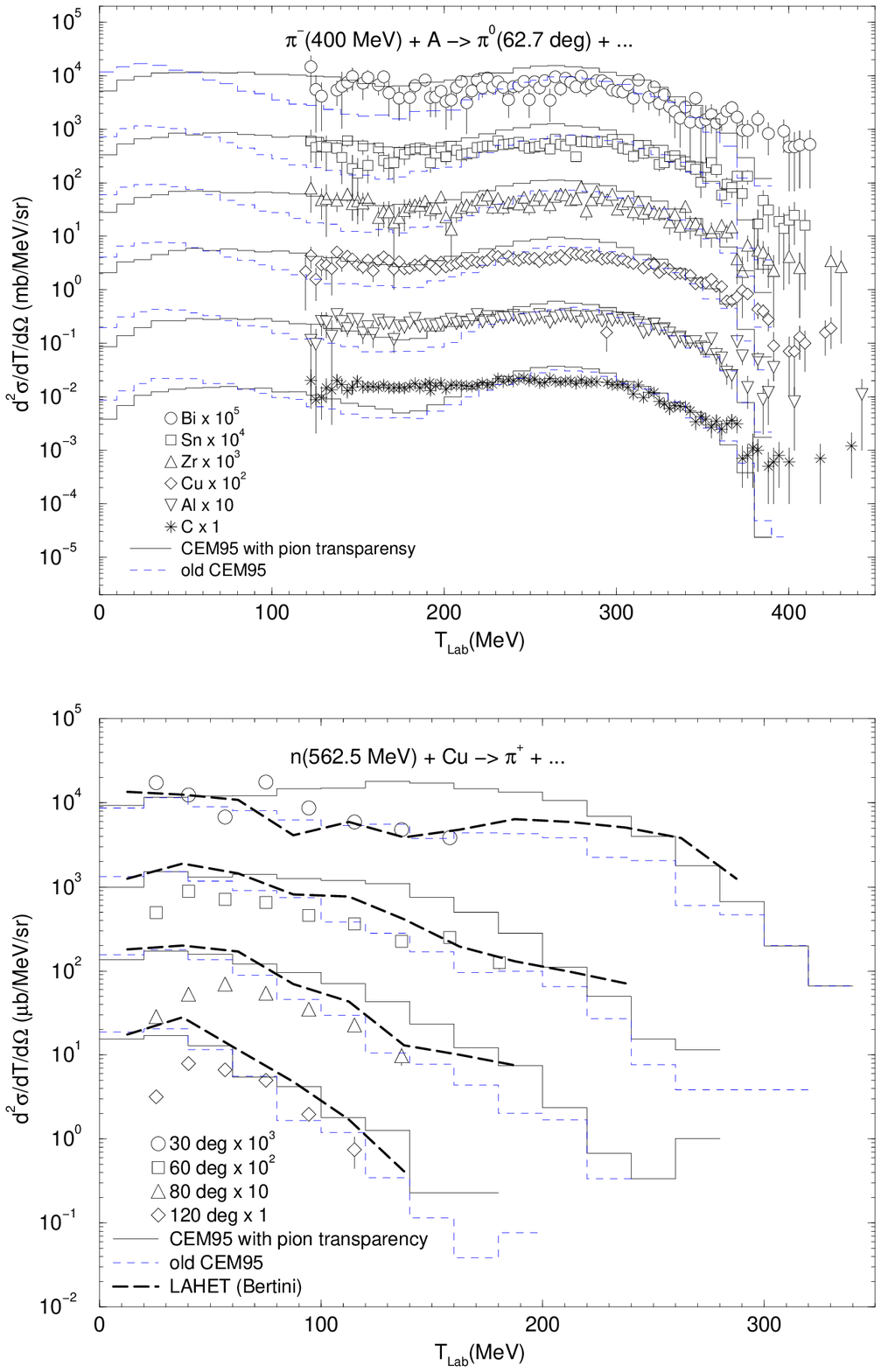,width=125mm,angle=0}}
\end{figure}
\vspace*{-0cm}

{\small
{\bf Fig.~8}.
Comparison of experimental SCX \cite{hoibrten} $\pi^0$ spectra 
from 400 MeV $\pi^-$ on different targets \cite{ouyang} and of $\pi^+$
spectra from 562.5 MeV neutrons on Cu \cite{brooks} with results of 
CEM95 calculations with pion transparency, i.e., with an imposed 
$\lambda_{\pi} \ge 2$ fm for pions from  $(\pi,2\pi)$ interactions
and $\lambda_{\pi} \ge 1.2$ fm for all pions after the first $\pi N$ 
interaction (solid histograms) and with standard CEM95 results 
(dashed histograms). For comparison, the thick dashed lines show 
results of LAHET (Bertini INC) \cite{lahet} calculations from \cite{brooks}.
\\
}

\noindent{
following absorption ratios \cite{engel}:
\begin{eqnarray}
\frac{\sigma_{\pi^+}(nn) \to np}{\sigma_{\pi^+}(np) \to pp} & = & 0.083,
 \quad 
\frac{\sigma_{\pi^0}(np) \to np}{\sigma_{\pi^+}(np) \to pp} = 0.440,
 \quad 
\frac{\sigma_{\pi^0}(nn) \to nn}{\sigma_{\pi^+}(np) \to pp} = 0.140
\mbox{ ,} \nonumber \\
\frac{\sigma_{\pi^0}(pp) \to pp}{\sigma_{\pi^+}(np) \to pp} & = & 0.140,
 \quad 
\frac{\sigma_{\pi^-}(pp) \to np}{\sigma_{\pi^+}(np) \to pp} = 0.083,
 \quad 
\frac{\sigma_{\pi^-}(np) \to nn}{\sigma_{\pi^+}(np) \to pp} = 1.
\end{eqnarray}

To test this approximation, we started with the compilation of 
all presently available experimental \\
cross sections for pion absorption on deuterons.
}
Then, we estimate ${\sigma_{\pi^-}(np) \to nn}$ using Eq.~(4) 
with $W = 4$. All other absorption cross sections, for other types 
of nucleon pairs and/or charges of pions are calculated according to 
Eq.~(5).   With such modifications, we study several characteristics
of nuclear reactions determined mainly by  pion absorption on nucleon pairs. 

Pion absorption makes a large contribution to spectra of secondary 
nucleons emitted at very backward angles in the energy region around 
140 MeV (see, e.g., \cite{cemnuclons}). Our first results, for backward 
proton spectra from proton-induced reactions, are very encouraging 
(see the upper plot in Fig.~9); using the ratios (5) for pion absorption
significantly improves the agreement of calculated spectra with the data.
But when we try these ratios for neutron spectra we get, as in the case
of pion transparency, an opposite result; neutron yields from the
pion absorption mode is strongly overestimated when using the ratios (5).
We note that a variation of the absolute normalization of the pion absorption
cross section, i.e., of $W$ would not solve this problem, since if we use
a smaller value for $W$ to adjust the neutron yields, then the proton
yields from pion absorption will be too low.  This negative result 
leads us to reject the ratios (5) for CEM and to return to equal probabilities
of pion absorption on different nuclear pairs.

In Fig.~9, both proton and neutron spectra at backward angles 
calculated with CEM95 (i.e., using equal probabilities for pion
absorption on different pairs) lie above the data by a factor of 2 at 
energies around 100 MeV. These nucleons come in our model
mainly from pion absorption. This means that pion absorption
cross sections used for these targets were too high and    $W$ 
in Eq.~(4) should be smaller. Similar results were obtained  
with  CEM95 for other light and medium targets.
To improve this situation, we re-examined the value of $W$. Fortunately,
we presently have reliable spectra of both protons and neutrons
measured at backward angles for the same targets by the same
experimentalists at energies around 100--200 MeV \cite{vikhrov}. 
In Fig.~10, we show part of our results: experimental 
spectra \cite{vikhrov} of neutrons and protons emitted at 140$^{\circ}$
from the interaction of 1 GeV protons with different nuclei are compared
with standard CEM95 calculations and using twice lower pion
absorption cross sections, i.e., $W = 2$ for all targets.
One can see that for medium and light nuclei, the
results obtained using $W=2$ agree much better with data as compared with
the standard CEM95 results ($W=4$). From these and similar
results obtained for other targets at other intermediate energies
we can conclude that $W$ seems to depend on the atomic number of the 
target; increasing from about 2 for light targets like C to about 4 for heavy
targets like Pb. Fig.~11, where we show results obtained with
$W=2$ for C, $W=3$ for Fe and Ni, and $W=4$ for Pb,
serves as a confirmation of this conclusion:
Introduction of a pion absorption cross section increasing with $A$ 
on nuclear pairs significantly improves the agreement with data
both for neutron and proton spectra at backward angles around 100--200 MeV. 
Similar results were obtained for other reactions. We have not yet fixed
the value of $W$ in our new version of the code, as we wish
to perform several more test calculations to determine $W$ more accurately. 
But we already see that we will need to use an $A$-dependent function for $W$, 
which increases from about 2 for C to about 4 for Pb. \\

\newpage

\begin{figure}[h!]
\vspace*{0.8cm}
\hspace*{2.5cm}
\psfig{figure=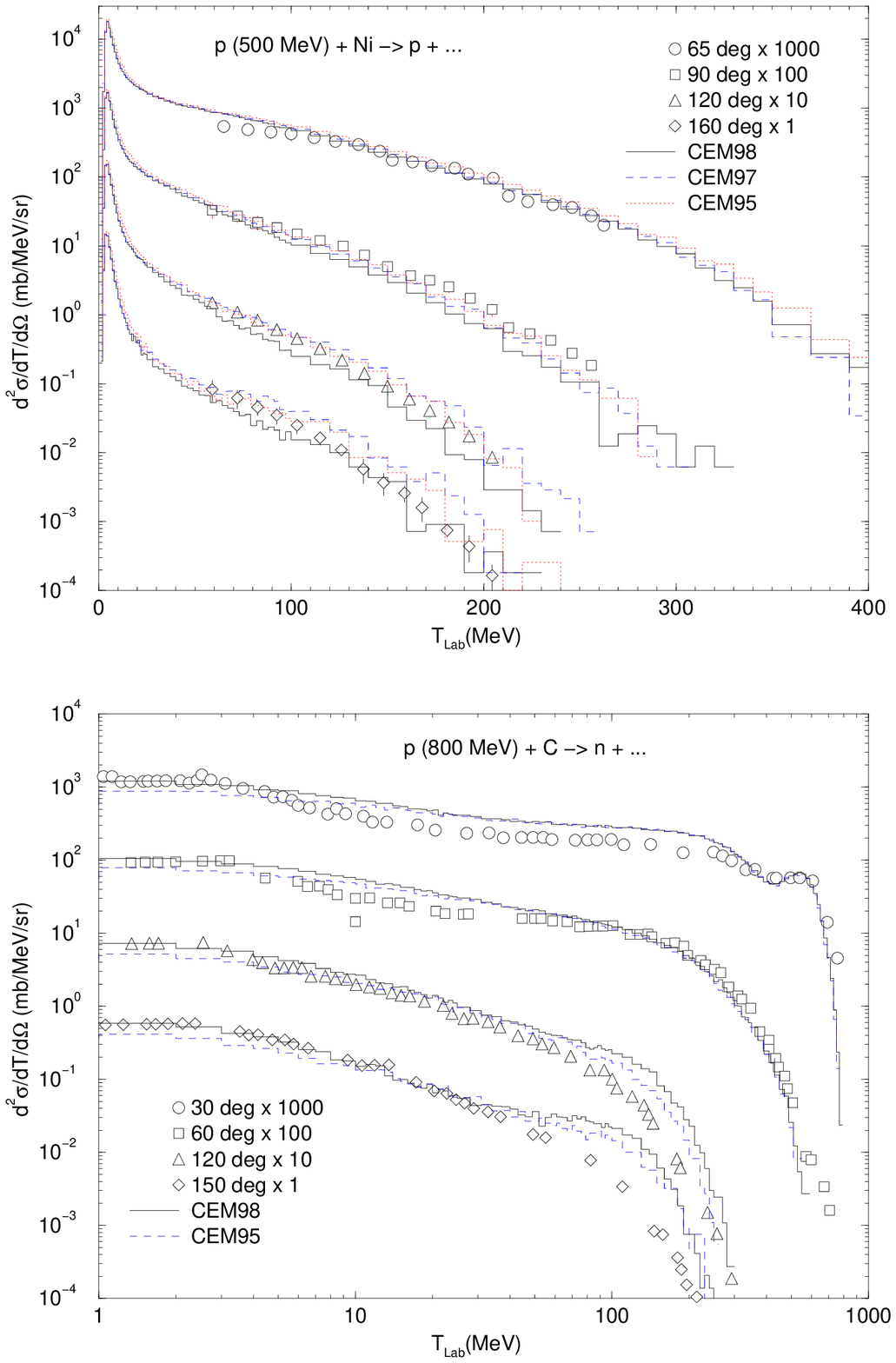,width=125mm,angle=0}
\end{figure}
\vspace*{0cm}

{\small
{\bf Fig.~9}.
Experimental proton spectra from 500 MeV p + Ni \cite{roy81}
and neutron spectra from 800 MeV p + C \cite{amian92}
compared with standard CEM95 results and with modified calculations 
of pion absorpion cross sections according to \cite{engel} (noted in this figure
as CEM98) as described in the text. CEM97 results are obtained with the same
pion abosrption cross sections as the CEM95 ones, but with new elementary
$NN$ and $\pi N$ cross sections and new nuclear masses.
}

\newpage

\begin{figure}[h!]
\vspace*{-3cm}
\centerline{
\psfig{figure=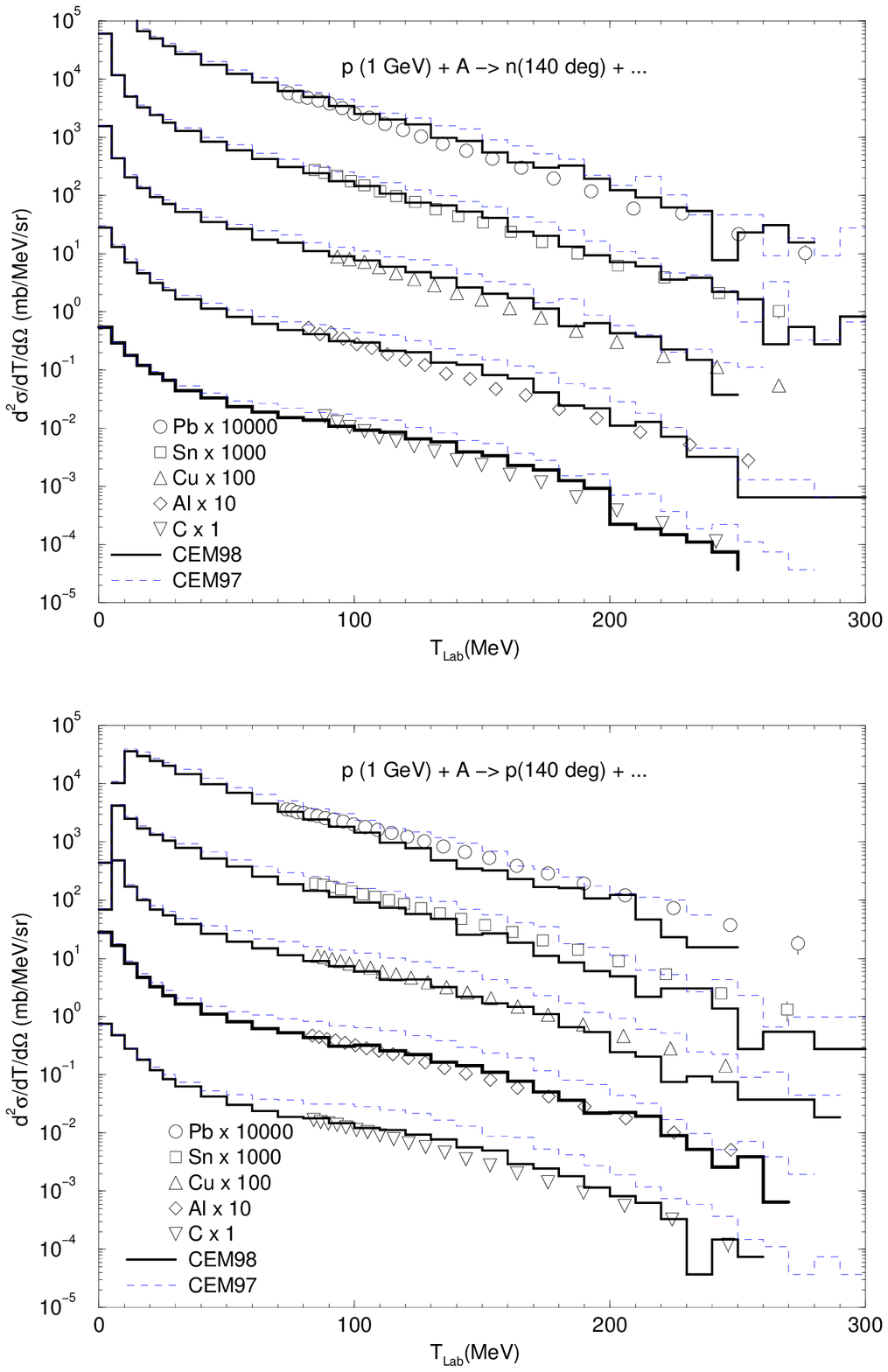,width=190mm,angle=0}}
\end{figure}
\vspace*{-4.5cm}

{\small
{\bf Fig.~10}.
Experimental spectra of neutrons and protons emitted at 140 degrees
from interactions of 1.0 GeV protons with C, Al, Cu, Sn, and Pb \cite{vikhrov}
compared with standard CEM97 results (dashed histograms)
and with calculations using one-half the pion absortpion cross sections, i.e., $W = 2$, or:
$\sigma_{abs}(\pi+[NN]) = 2 \sigma_{abs}(\pi+d])$ (noted here as CEM98).
\\
}

{\bf F. Other Enhancements, CEM97a Code, and Further Work}.
We have made a number of other improvements to CEM95
concerning the description of the preequilibrium and evaporation
stages of reactions. This includes incorporating a 
complete experimental mass table extended by the
Moller-Nix mass calculations \cite{molnix}, along with corresponding ground-state 
microscopic corrections and pairing energies.
We have also implemented a new level-density approximation which 
incorporates the calculated microscopic corrections and pairing energies
and angular momentum dependent macroscopic fission 
barriers~\cite{sierk86}, 

\newpage

\begin{figure}[h!]
\vspace*{-3cm}
\centerline{
\psfig{figure=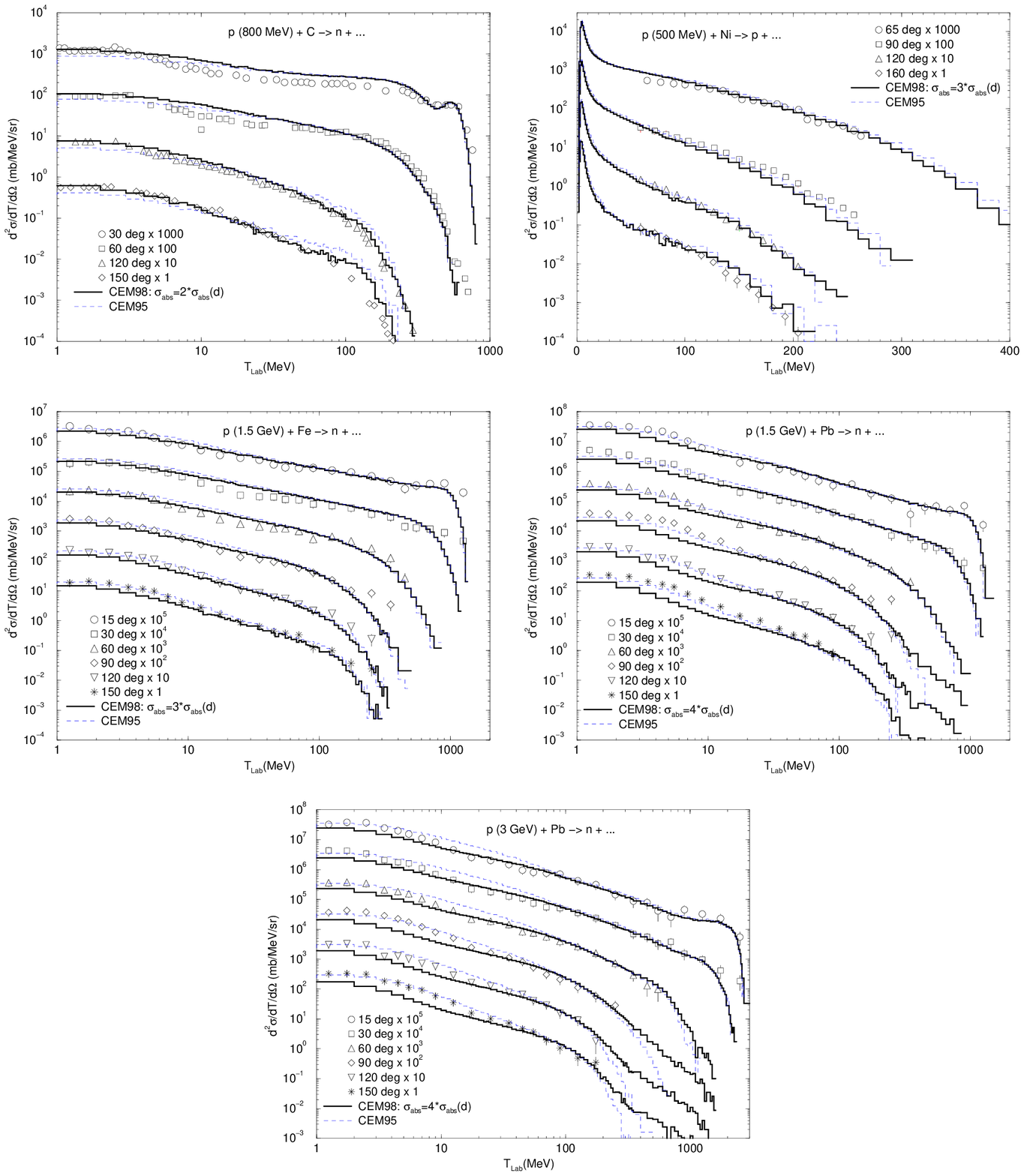,width=198mm,angle=0}}
\end{figure}
\vspace*{-3.5cm}

{\small
{\bf Fig.~11}.
Comparison of  experimental neutron (proton, for 500 MeV protons on Ni)
spectra from interactions of protons of 800 MeV on C \cite{amian92},
500 MeV on Ni \cite{roy81}, 1.5 GeV on Fe and Pb \cite{ishibashi97},
and 3.0 GeV on Pb \cite{ishibashi97}, with standard CEM95 results 
(dashed histograms) and with calculations using modified 
pion absorption cross sections on quasideuterons, namely,
$W = 2$ for the light C target, $W = 3$ for the medium mass Ni and Fe 
targets, and $W = 4$ for the heavy Pb target (noted here as CEM98).
\\
}

\newpage

\noindent{
which are consistent with the macroscopic model used in 
the Moller-Nix mass calculations.
We have significantly improved the treatment of rotational energy and have made 
a number of small refinements to
the code like using actual updated values for the masses of elementary 
particles (for simplicity in CEM95, we used, e.g.,
$m_\pi = 140$ MeV for pions of any charges and $m_n = m_p = 940$ MeV)
and more precise values for physical constants.

With these modifications and various coding improvements, we wrote a 
preliminary improved version of the CEM95 code called CEM97a, 
which is now incorporated in the MCNPX transport code \cite{inmcnpx}. 

CEM97a is a preliminary release, since several improvements have been
made after its creation and we continue the work in this
direction, but already CEM97a provides a much better description
of many data than the initial code, CEM95. 
}
As an example, in Fig.~12 we show excitation functions for production of all
Xe isotopes for which we found experimental data from p+$^{133}$Cs interactions, 
 part of our calculations for a medical isotope production study \cite{medical}, 
using the standard CEM95 and with the improved version CEM97a. 
All available experimental data and calculations with the LAHET \cite{lahet} 
code system (version 2.83) by K.~A.~Van Riper from \cite{medical}
and with the HMS-ALICE code \cite{alice96} by M.~B.~Chadwick from the 
LA150 Activation Library \cite{mark} are shown as well, for comparison.
One can see that CEM97a describes these data much better than
CEM95 and LAHET, and the difference between predictions
of the new and old versions of CEM increases as the isotopes produced become
more neutron deficient, reaching a factor of 3 for $^{120}$Xe. 
Such a big difference between results from the new and old versions 
of the code is related mainly to using experimental nuclear 
masses and binding energies in CEM97a and the 40 year old approximations
of Cameron \cite{cam57} in CEM95.

An example of the importance of using reliable values for nuclear masses
can be seen also in Fig.~11. From this figure one can see that neutron
spectra from Pb calculated with CEM95 
agree very well with the data at low energies, in the evaporation parts
of spectra. This is not necessarily good, since the calculations with CEM95 
for Pb take into account competition between evaporation and fission, but 
the fission itself is not calculated; when a fission event occurs
after the cascade and preequilibrium stages of a reaction, CEM95 stores 
this event to calculate a fission cross section, but then it stops further
simulation of this event. That is, calculated spectra do not contain
contributions of particles evaporated from fragments following fission.
We are presently working to incorporate simulation of the fission
processes themselves into the CEM, and we need some ``room" for
particles evaporated from fission fragments. As we see from Fig.~11,
the new version (noted as CEM98) of the code provides us such ``room";
calculated spectra agree very well with the data for intermediate and
high energy parts of the spectra but lie below the measurements in the
evaporation region by a factor of 2, leaving some place for neutrons
evaporated from fission fragments.

After creating CEM97a, we have further improved the code. We have
improved the approximation to the level densities 
at the fission saddle point and have introduced other modifications,
given separately in \cite{modfiss}. We also further modified
the cascade modeling in the code by improving the treatment of 
photo-nuclear reactions and imposing energy and momentum conservation 
for the whole cascade stage of a reaction using experimental nuclear 
masses and binding energies for incident and emitted nucleons
instead of the old approximations of CEM95.

Our improvement efforts continue. We have written a subroutine for 
emission of complex particles at the cascade stage of reactions via the
mechanism of their coalesence from emitted nucleons, have modified the
preequilibrium and evaporation stages to allow emission of fragments
with $A > 4$ from not too light excited nuclei, and are working now
to develop an appropriate model for high energy fission. Among many
other things, we need to improve the approximation of the inverse 
cross sections used in calculating particle emission widths and to treat 
more accurately $\alpha$-emission at both the preequilibrium and 
evaporation stages.

\newpage

\begin{figure}[h!]
\vspace*{-0.2cm}
\centerline{
\psfig{figure=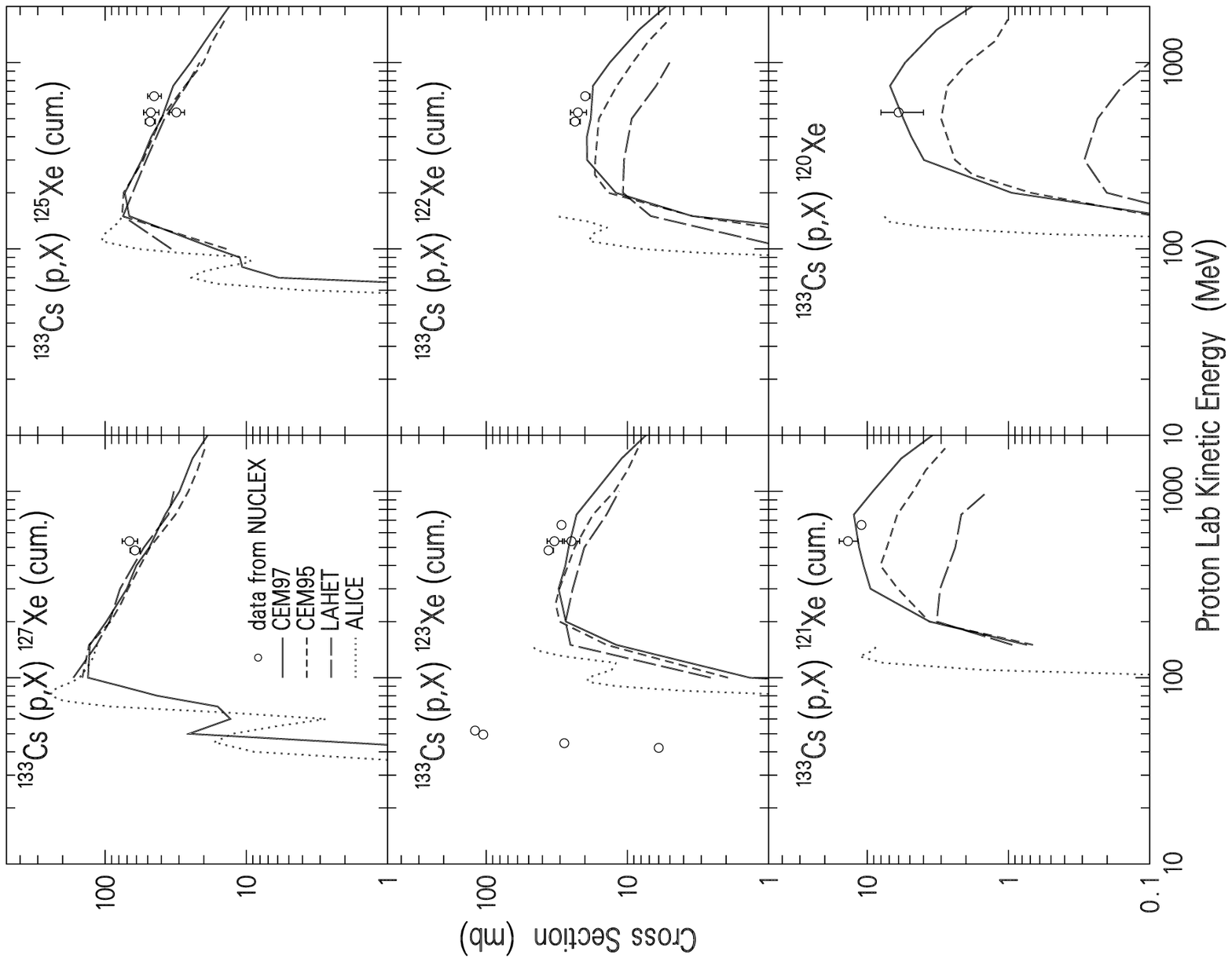,width=250mm,angle=-90}}
\end{figure}
\vspace*{+0.9cm}

{\small
{\bf Fig.~12}.
Excitation functions for the production of $^{127-120}$Xe from
p+$^{133}$Cs. Calculations with the new CEM97a code
are shown by solid lines and with the standard CEM95 by short dashed
lines.  Results of calculations with the LAHET \cite{lahet} code system
(version 2.83) by K.~A.~Van Riper from Ref. \cite{medical} are shown by
long dashed lines and results of the HMS-ALICE code \cite{alice96} 
calculated by M.~B.~Chadwick from the LA150 Activation Library \cite{mark}
are shown by dotted lines. Experimental data are from the compilation 
\cite{nuclex}.
\\
}

\begin{center}
{\large 4. Summary} \\
\end{center}

In this paper, we have demonstrated the good overall predictive power
of the modified version of the CEM as realized in the code CEM95.
Then, to improve the agreement of its results with experimental
data and to make it a better tool for applications, we 
modify it further, progressively incorporating features of previously 
neglected physics. So far, we have incorporated in the CEM new and 
better approximations for the elementary cross sections, 
imposed momentum-energy conservation for each simulated event, 
used more precise values for nuclear masses, $Q$-values, binding and pairing
energies, corrected systematics for the level density parameters,
studied and chosen the optimal approximations for the pion
``binding energy", $V_{\pi}$, for the cross sections of pion
absorption on quasideuteron pairs inside a nucleus, for the effects of
refractions and reflections, and for nuclear transparency of pions.
We also make a number of refinements in calculation of the
fission channel, described separately in \cite{modfiss}.

As it was shown by a number of examples,
the improvements to the CEM made so far clearly have increased its
predictive power. Our work is not finished. Among improvements of the
CEM which are of highest priority we consider development and incorporation
of an appropriate model of high-energy fission, treating more accurately 
$\alpha$-emission at both preequilibrium and evaporative stages,
incorporation of a model of fragmentation of medium and heavy nuclei,
possibly the Fermi breakup model for highly excited light nuclei,
modeling evaporation and preequilibrium emission of fragments with
$A > 4$, and improvement of the approximation for inverse cross sections.

The problems discussed above are typical not only for the CEM, but also
for all other similar models and codes, where they are also not yet solved.
\begin{center}
{\it Acknowledgements } 
\end{center}
We express our gratitude to K.~Ishibashi, S.~Chiba, J.~D.~Zumbro, 
R.~J.~Peterson, V.~V.~Vikhrov,
and B.~Bassaleck for suppling us with numerical values of their measurements
and we thank K.~Ukai for kindly providing us with his and T.~Nakamura's 1997 
compilation {\em ``Data Compilation of Single Pion Photoproduction Below 2 
GeV"} used in our work.
We are grateful to R.~E.~MacFarlane and L.~S.~Waters for helpful discussions 
and support of the present work. We thank many users of the CEM codes,
especially V.~F.~Batyaev, O.~Bersillon, F.~Gallmeier, N.~V.~Mokhov,  and
A.~V.~Prokofiev, for their constructive contributions,
which led to the removal of several bugs and to improvements in the model.
This study was supported by the U.~S.~Department of Energy.

\end{document}